\documentstyle[11pt,epsfig]{article}

\def\laq{~\raise 0.4ex\hbox{$<$}\kern -0.8em\lower 0.62
ex\hbox{$\sim$}~}
\def\gaq{~\raise 0.4ex\hbox{$>$}\kern -0.7em\lower 0.62
ex\hbox{$\sim$}~}
\begin{document}

\begin{titlepage}

\begin{flushright}
CERN-PH-TH/2004-128
\end{flushright}

\vspace*{1.8 cm}

\begin{center}

\huge
{Vector fluctuations \\ from multidimensional curvature bounces}

\vspace{1cm}

\large{Massimo Giovannini}

\normalsize
\vspace{.1in}

{\sl Department of Physics, Theory Division, CERN, 1211 Geneva 23, Switzerland }

\vspace*{2.5cm}
\begin{abstract}
It is argued that in the case of a smooth transition across a (dilaton-driven) curvature bounce the growing mode of the 
vector fluctuations matches continuously with a decaying mode at later times.  Analytical examples 
of this observation are given both in the presence and in the absence of fluid sources. 
In the case of multidimensional bouncing models the situation is different, since the system of differential equations 
describing the vector modes of the geometry has a richer structure. The amplification of the vector modes 
of the geometry  is specifically investigated in a regular five-dimensional bouncing curvature model where 
 scale factors of the external and internal manifolds evolve at a dual rate. Vector fluctuations, in this case, 
can be copiously produced and are continuous across the bounce. The relevance of these 
results is critically illustrated.
\end{abstract}
\end{center}
\end{titlepage}
\newpage

\renewcommand{\theequation}{1.\arabic{equation}}
\section{Formulation of the problem}
\setcounter{equation}{0}

It is known that in conventional inflationary models vector fluctuations of the metric
are not amplified by the pumping action of the gravitational field \cite{gr1}.
On the contrary both scalar and tensor modes of the geometry are copiously 
generated in the transition from a quasi-de Sitter stage of expansion 
to a radiation-dominated evolution.

Vector modes of the geometry may have interesting 
phenomenological implications. For instance, as argued long ago by Wassermann \cite{wass} (see also 
\cite{coles}) the existence of large-scale vorticity (sometimes connected with a large-scale magnetic field) 
may have relevant implications for the origin of global rotation of spiral galaxies. 
The possible existence of primordial  vorticity has been the subject of various investigations. 
Harrison \cite{harrison3} (see also \cite{mishustin} for a complementary perspective) 
noticed that if vorticity is present over sufficiently 
large length scales in a plasma of ions and electrons,  magnetic fields can be 
induced over scales comparable with the typical scale of the vorticity. The dynamical origin of primordial 
vorticity has been, since then, a subject of debate. In \cite{vilenkin1} it was suggested that 
cosmic strings with small scale structure may be responsible for the dynamical origin of vorticity. 
The dynamical friction between cosmic strings and matter may also be responsible 
for further sources of vorticity \cite{shellard}. 

In spite of the fact that the considerations quoted so far may indeed produce large-scale vorticity, it 
is useful to take another approach and consider the possibility that, thanks to the pumping action of the gravitational field, 
the vector fluctuations of the geometry will be amplified, providing, at some later epoch, the 
initial condition for the evolution of the rotational fluctuations of the fluid.
The motivation of the present investigation is to discuss the evolution of vector modes of the geometry 
in the case of a bouncing background. As is known,  the ten-dimensional 
degrees of freedom of a four-dimensional metric can be decomposed, with respect to three-dimensional 
rotations,  into four scalar modes, two tensor modes (described by a divergenceless and  traceless 
rank-two symmetric tensor in three dimensions), and four vector degrees of freedom (described by two divergenceless 
vectors in three dimensions). Out of these ten degrees of freedom, four can be fixed by 
a choice of coordinate system. While the tensor modes are invariant under infinitesimal gauge transformations, the scalar and vector modes are not.  

While this work was in progress, Battefeld and Brandenberger \cite{BB} argued that it is 
not unlikely that vector modes of the geometry may be amplified in the case of four-dimensional 
curvature bounces as the ones provided by cyclic/ekpyrotic scenarios \cite{ekp} or in the context 
of the  pre-big bang scenario \cite{pbb}.  The authors of Ref. \cite{BB} worked in a four-dimensional 
conformally flat FRW  model and discussed the dynamics in the context of an Einsteinian theory of gravity. 
They found a growing-vector zero mode and correctly 
pointed out that the specific fate of  such a mode  is rather sensitive to the 
detailed dynamics of the bounce. 

There are, in principle, different bouncing dynamics that can be analysed. The simplest case 
is the one of a four-dimensional background without fluid sources but with a scalar 
degree of freedom (the dilaton). In this 
case, during the contraction the vector zero mode increases, but as  will be argued in specific examples, its fate depends upon the dynamics
 of the curvature bounce and, in all the 
analytical examples reported in the present paper, it matches with a decaying mode after the 
bounce. A similar situation also arises, also in four dimensions but in the presence of fluid sources
and together with a dilatonic background. 

In the multidimensional case the situation becomes increasingly interesting. 
There more scalar and vector modes arise  and their 
quantum mechanical fluctuations can be amplified. 

For the sake of completeness, even if the scalar modes of the geometry are not the subject of the present 
investigation, we want to mention that
curvature bounces have been the subject of various debates concerning the fate of the 
scalar fluctuations of the geometry (see \cite{ref} for a partial list of references).  

The plan of the present paper will be  the following. In Section 2 the evolution 
of vector modes of the geometry will be analysed in the context of four-dimensional curvature bounces.
In the string frame description, it will be shown that the vector modes present prior to the 
curvature bounce match continuously with a decreasing vector mode after the bounce. Specific 
analytical examples will be presented both in the absence and in the presence of fluid sources.
In Section 3 the specific case of multidimensional curvature bounces will be discussed. 
The evolution of the vector fluctuations of a five-dimensional curvature 
bounce will be derived in fully gauge-invariant terms. In Section 4 the spectrum 
of the vector modes will be derived and it will be shown that vector modes can be 
amplified in the five-dimensional model of curvature bounce. Section 5 contains 
the concluding remarks with a critical summary of the main suggestions stemming from 
the present study. Finally, the appendix contains useful technical results needed 
for the derivation of the evolution equations of vector modes in more than four dimensions.

\renewcommand{\theequation}{2.\arabic{equation}}
\section{Four-dimensional curvature bounces}
\setcounter{equation}{0}

Consider, to begin with, the low-energy gravi-dilaton action in four space-time dimensions 
and in the string frame description \cite{st}:
\begin{equation}
S_{4} = - \frac{1}{2 \lambda_{s}^2} \int d^4 x \sqrt{- G} e^{- \varphi} \biggl[ 
R + G^{\alpha\beta} \partial_{\alpha}\varphi \partial_{\beta} \varphi  + V (\varphi) \biggr].
\label{locac}
\end{equation}
The potential $V(\varphi)$ is, for the moment, a local function 
of the dilaton field $\varphi$ and $\lambda_{s}$ is the string length.
 A linear combination of the equations of motion 
derived from  Eq. (\ref{locac}) leads to the well known form of the low-energy beta functions
\begin{equation}
R_{\mu\nu} + \nabla_{\mu} \nabla_{\nu} \varphi - \frac{1}{2} g_{\mu\nu} \frac{\partial V}{\partial\varphi}  =0.
\label{beta}
\end{equation}
In a conformally flat FRW background and assuming the dilaton field is  homogeneous  
\begin{equation}
ds^2 = G_{\alpha\beta}  dx^{\alpha} d x^{\beta}=
 a^2(\eta) [d\eta^2 - d\vec{x}^2],\,\,\,\,\,\,\,\,\,\varphi= \varphi(\eta),
\end{equation}
the $(i,j)$ component of Eq. (\ref{beta}) leads to the following relation
\begin{equation}
{\cal H}' - {\cal H}^2 - {\cal H} \overline{\varphi}' + \frac{a^2}{2} \frac{\partial V}{\partial \varphi} =0.
\label{b0}
\end{equation}
In Eq. (\ref{b0}) the 
prime denotes a derivation with respect to the conformal time coordinate $\eta$ while ${\cal H} 
= (\ln{a})'$ and   
$\overline{\varphi}' = \varphi' - 3{\cal H}$ is the time  derivative of the so-called shifted dilaton, which is
invariant under scale-factor duality.

In this framework, the vector modes of the geometry are parametrized by two vectors, $Q_{i}$ and $W_{i}$, which 
are divergenceless, i.e. $\partial_{i}Q^{i}= \partial_{i} W^{i} =0$ \footnote{In the present investigation 
the attention will be confined to spatially flat manifolds. However, our results can be 
extended to the case of non-flat spatial geometry by replacing, when needed, the ordinary derivatives 
with covariant derivatives with respect to the spatial part of the geometry.}:
\begin{eqnarray}
\delta G_{0i} & =& - a^2(\eta) Q_{i},
\nonumber\\
\delta G_{ij} &=& 2 a^2(\eta)  \partial_{(i} W_{j)} \equiv a^2(\eta) \biggl( \partial_{i} W_{j} + \partial_{j}
W_{i} \biggr).
\label{fluc}
\end{eqnarray}
From Eq. (\ref{fluc}),  the components Ricci tensor can  be  computed to first order in the fluctuations of the 
metric : 
\begin{eqnarray}
&& \delta R_{0i} = ({\cal H}' + 2 {\cal H}^2) Q_{i} - \frac{1}{2} \nabla^2[ Q_{i} + W_{i}'] ,
\nonumber\\
&& \delta R_{ij} = - \partial_{(i} Q_{j)}' - \partial_{(i} W_{j)}''  - 2 ({\cal H}' + 2 {\cal H}^2) \partial_{(i} W_{j)}
- 2 {\cal H} [ \partial_{(i} W_{j)}' + \partial_{(i} Q_{j)}].
\label{ricci1} 
\end{eqnarray}

Finally, using Eq. (\ref{ricci1}), the first-order fluctuation of  Eq. (\ref{beta})  leads to the following pair of differential equations:
\begin{eqnarray}
&&
\biggl[{\cal H}' - {\cal H}^2 - {\cal H} \overline{\varphi}' +\frac{ a^2}{2} \frac{\partial V}{\partial \varphi}\biggr] Q_{i} - 
\frac{1}{2} \nabla^2( Q_{i} + W_{i}') =0, 
\label{int0}\\
&& 
-   \partial_{(i} Q_{j)}' - \partial_{(i} W_{j)}''   + (\overline{\varphi}' + {\cal H}) \partial_{(i} Q_{j)} + \partial_{(i} W_{j)}'  
\nonumber\\
&&- 2 \biggl[{\cal H}' - {\cal H}^2 - {\cal H} \overline{\varphi}' +\frac{ a^2}{2} \frac{\partial V}{\partial \varphi}\biggr]
\partial_{(i} W_{j)} =0.
\label{int1}
\end{eqnarray}
Recalling the specific form of Eq. (\ref{b0}), it can be appreciated that  various terms cancel in Eqs. (\ref{int0}) and (\ref{int1}).
Defining now the  new variable
\begin{equation}
V_{i} = Q_{i} + W_{i}', 
\label{Vi}
\end{equation}
Eqs. (\ref{int0}) and (\ref{int1}) can be written as 
\begin{eqnarray}
&&\nabla^2 V_{i} =0,
\label{v01}\\
&& V_{i}' = (\overline{\varphi}' + {\cal H}) V_{i} .
\label{v02}
\end{eqnarray}

The variable $V_{i}$ is invariant under infinitesimal coordinate transformations, preserving the vector nature of the fluctuation.
Indeed, under infinitesimal diffeomorphisms $Q_{i}$ and $W_{i}$ transform as
\begin{eqnarray}
&&\tilde{W}_{i} = W_{i}  +  \zeta_{i},
\nonumber\\
&& \tilde{Q}_{i} = Q_{i}  -  \zeta_{i}',
\end{eqnarray}
where $\partial_{i} \zeta^{i} =0$ is the gauge function. Hence, as anticipated, $V_{i}$ 
is invariant under infinitesimal coordinate transformations. This observation 
also implies that the condition  $W_{i}=0$ fixes the gauge freedom completely and is analogous, in this 
sense, to the conformally Newtonian gauge often employed in the analysis of scalar perturbations \footnote{Notice that, in the scalar 
case, the longitudinal gauge condition (imposed on the gauge parameters preserving the scalar 
nature of the fluctuation) eliminates the off-diagonal entries of the perturbed metric. In the case of vector modes the situation is, in a sense, opposite since
the gauge choice fixing  the coordinate system completely keeps only the off-diagonal elements of the perturbed metric tensor.}.
This gauge  choice would have led to the evolution equations for $Q_{i}$, coinciding,
in such a gauge, with the evolution of the gauge-invariant vector $V_{i}$. We will go back to the gauge issue 
during the discussion of vector fluctuations in the multidimensional case.

From Eqs. (\ref{v01}) and (\ref{v02}) we can deduce that only the 
zero mode is dynamical. In this case its evolution can be written as
\begin{equation}
V_{i} = c_{i}\,\, a(\eta)\,\, e^{\overline{\varphi}},
\label{sol0}
\end{equation}
where $c_i$ is an integration constant. During a conventional pre-big bang phase \cite{rev}, 
as $\eta \to 0^{-}$, the consistent solution of Eq. (\ref{b0}) and of the other 
equations of the background can be written, in the absence of dilaton potential, as \cite{rev}
\begin{eqnarray}
&& a(\eta) \sim (-\eta)^{- \frac{1}{\sqrt{3} + 1}}\sim  (- t)^{- \frac{1}{\sqrt{3}}},
\nonumber\\
&& \overline{\varphi} \sim - \frac{\sqrt{3}}{\sqrt{3} +1} \ln{|\eta|}\sim - \ln{|t|},
\label{solb1}
\end{eqnarray}
 where $ t$ is the cosmic time coordinate obeying the usual differential relation $a(\eta) d\eta = dt$.
Inserting Eq. (\ref{solb1}) into Eq. (\ref{sol0}) 
\begin{equation}
 V_i \sim |\eta|^{-1} \sim |t|^{- \frac{\sqrt{3} + 1}{\sqrt{3}}}.
 \end{equation}
This growing mode diverges in the limit $t \to 0^{-}$.
 This is essentially one of the arguments put forward in Ref. \cite{BB}, with the 
 difference that, in the present case, the discussion has been conducted, from the 
 very beginning, in the string frame description.   

The aim  will now be to show that in a simple model of low-curvature transition, the growing vector mode will 
match to a decreasing solution in the post-bounce epoch.  As was  shown in 
 \cite{pbb} a smooth transition at low curvatures can be achieved if the dilaton 
potential is a local function of $\overline{\varphi}$.  The generally covariant action 
describing this model is given, in $D$ space-time dimensions by 
\begin{equation}
S= - \frac{1}{\lambda_{s}^{D-2} }\int d^{D-1} x \sqrt{|G|} e^{- \varphi} \biggl[ R + 
G^{\alpha\beta} \nabla_{\alpha} \varphi \nabla_{\beta} \varphi + V(\overline{ \varphi})\biggr] + S_{\rm m},
\label{action1}
\end{equation}
where $S_{\rm m}$ represents the contribution of effective (fluid) sources and where 
\begin{eqnarray}
&& V= V(e^{-\overline{\varphi}}),
\nonumber\\
&& e^{-\overline{\varphi}(x)} = \frac{1}{\lambda_{s}^{D-1}} \int d^{D} w \sqrt{|G(w)|} e^{- \varphi(w)} 
\sqrt{G^{\alpha\beta} \partial_{\alpha} \varphi (w)\partial_{\beta} \varphi(w)} \delta\biggl( \varphi(x) - \varphi(w)\biggr).
\end{eqnarray}
The variation of the action (\ref{action1}) with respect to $G_{\mu\nu}$ and $ \varphi$ leads to the following equations:
\begin{eqnarray}
&&{\cal G}_{\mu\nu}  + \nabla_{\mu} \nabla_{\nu} \varphi + 
\frac{1}{2} g_{\mu\nu} \biggl[ G^{\alpha\beta} \nabla_{\alpha}\varphi \nabla_{\beta} \varphi - 2 
G^{\alpha\beta} \nabla_{\alpha} \nabla_{\beta} \varphi - V\biggr], 
\nonumber\\
&& - \frac{e^{-\varphi}}{2} 
\sqrt{ G^{\alpha\beta} \partial_{\alpha}\varphi \partial_{\beta} \varphi} \gamma_{\mu\nu} {\cal I}_{1} = 
\lambda_{s}^{D -2} T_{\mu\nu},
\label{eq1}\\
&& R + 2 G^{\alpha\beta} \nabla_{\alpha} \nabla_{\beta} \varphi - 
G^{\alpha\beta} \partial_{\alpha} \varphi\partial_{\beta} \varphi + V - \frac{\partial V}{\partial\overline{\varphi}} 
\nonumber\\
&& + e^{-\varphi} \frac{\hat{\nabla}^2 \varphi}{\sqrt{G^{\alpha\beta} \partial_{\alpha}\varphi\partial_{\beta}\varphi}} {\cal I}_1 - e^{-\varphi} V' {\cal I}_{2} =0,
\label{eq2}
\end{eqnarray}
where \footnote{In Eq. (\ref{eq2}) as well as in  Eqs. (\ref{def}) and (\ref{comb}) the prime denotes a derivation with respect to the 
argument of the given functional and {\em not} the derivative with respect to the conformal time coordinate. The two notations cannot be confused since the 
latter is only employed when dealing with the explicit form of the equations on a given background.} 
\begin{equation}
\hat{\nabla}^2 = \gamma^{\mu\nu} \partial_{\mu}\partial_{\nu},\,\,\,\,\,\,\,\,\,\,\,\,\,\,{\cal G}_{\mu}^{\nu} = R_{\mu}^{\nu} - 
\frac{1}{2} \delta_{\mu}^{\nu} R, 
\end{equation}
and 
\begin{equation}
\gamma_{\mu\nu} = g_{\mu\nu} - \frac{\partial_{\mu} \varphi \partial_{\nu} \varphi}{\sqrt{G^{\alpha\beta} 
\partial_{\alpha}\varphi \partial_{\beta}\varphi}}
\label{induced}
\end{equation}
is the induced metric. In Eqs. (\ref{eq1}) and (\ref{eq2})  the following integrals 
\begin{eqnarray}
&& {\cal I}_1 = \frac{1}{\lambda_{s}^{D -1}} \int d^{D} w \sqrt{|G(w)|} V' (e^{- \overline{\varphi}(w)}) 
\delta(\varphi(x) - \varphi(w)),
\nonumber\\
&& {\cal I}_{2} = \frac{1}{\lambda_{s}^{D-1}} 
\int d^{D} w \sqrt{|G(w)|} \sqrt{G^{\alpha\beta} \partial_{\alpha}\varphi(w) \partial_{\beta} \varphi(w)}\delta'(\varphi(x) - \varphi(w)),
\label{def}
\end{eqnarray}
have also been defined.

By combining Eqs. (\ref{eq1}) and (\ref{eq2}) in order to eliminate the Ricci scalar 
we obtain the following equation
\begin{eqnarray}
&&R_{\mu\nu} + \nabla_{\mu}\nabla_{\nu} \varphi - \frac{1}{2} G_{\mu\nu} \biggl[ \frac{\partial V}{\partial 
\overline{\varphi}} + e^{-\varphi} V' {\cal I}_{2} \biggr] + \frac{1}{2} e^{-\varphi} \biggl[  G_{\mu\nu} 
\frac{\hat{\nabla}^2 \varphi}{\sqrt{G^{\alpha\beta} \partial_{\alpha}\varphi \partial_{\beta} \varphi}}
\nonumber\\
 && - 
\gamma_{\mu\nu} \sqrt{G^{\alpha\beta} \partial_{\alpha}\varphi\partial_{\beta}\varphi}\biggr] {\cal I}_{1} 
= \lambda_{s}^{D-2} T_{\mu\nu}.
\label{comb}
\end{eqnarray}
If  ${\cal I}_{1}$ and ${\cal I}_{2}$ are set to zero and if  $ \partial V/\partial \overline{\varphi}$ is replaced by
 $ \partial V/\partial {\varphi}$, Eq. (\ref{beta}) can be formally recovered from Eq. (\ref{comb}).

In the case of a homogeneous dilaton and for a conformally flat metric of FRW type Eqs. (\ref{eq1}) and (\ref{comb}) lead,  in $D=4$ and 
in units $2 \lambda_{s}^2 =1$, to the following system of equations: 
\begin{eqnarray}
&& \dot{\overline{\varphi}}^2 - 3 H^2 - V = e^{\overline{\varphi}} \overline{\rho},
\label{b1}\\
&& \dot{H} = \dot{\overline{\varphi}} H + \frac{1}{2} e^{\overline{\varphi}} \overline{p},
\label{b2}\\
&& 2 \ddot{\overline{\varphi}} - \dot{\overline{\varphi}}^2 - d H^2 + V - \frac{\partial V}{\partial \overline{\varphi}} =0,
\label{b3}\\
&& \dot{\overline{\rho}} + d H \overline{p} =0,
\label{b4}
\end{eqnarray}
 where the stress tensor of the fluid sources
\begin{equation}
T_{\mu}^{\nu} =( p+ \rho) u_{\mu}u^{\nu} - p \delta_{\mu}^{\nu} 
\end{equation} 
has been assumed and where the overdot denotes a derivation with respect to the cosmic time coordinate.
In Eqs. (\ref{b1})--(\ref{b3}) the energy and pressure densities have been rescaled as 
\begin{equation}
\overline{\rho} = a^{3} \rho, ~~~~~~~\overline{p} = a^{3} p.
\end{equation}
In the case $ \overline{p} =\overline{\rho}=0$,  Eqs. (\ref{b1})--(\ref{b4}) 
admit the solution \cite{pbb} 
\begin{eqnarray}
&& V (\overline{\varphi}) = - V_0 e^{ 4\overline{\varphi} },
\label{pot1}\\
&&a(t) = a_0 \biggl[ \tau + \sqrt{\tau^2 + 1}\biggr]^{1/\sqrt{3}},
\label{scalefactor1}\\
&& \overline{\varphi} = - \frac{1}{2} \log{ ( 1 + \tau^2)} + \varphi_{0},
\label{dilaton1}
\end{eqnarray}
where
\begin{equation}
\tau = \frac{t}{t_0}, ~~~~~~~~~~~~ t_0 = \frac{e ^{- 2 \varphi_0}}{ \sqrt{V_{0}}}.
\end{equation}
The scale $t_0$ fixes the width of the curvature bounce \footnote{The possibility of 
expressing the evolution of the background geometry in terms of the rescaled cosmic time $\tau$ 
will also prove useful in the numerical discussion of the spectrum of vector fluctuations in multidimensional models.}.

Another solution, which is effective in illustrating some aspects of the present analysis, may be obtained 
if the scale factor and the energy density of the fluid are both {\em constant}. In this toy example,
Eqs. (\ref{b1})--(\ref{b4}) may be integrated easily and the result is
\begin{eqnarray}
&& a= 1,\,\,\,\,\,\,\,\,\,\,\,\rho = \rho_0,\,\,\,\,\,\,p=0,
\nonumber\\
&& V = - V_0 e^{2\overline{\varphi}},\,\,\,\,\, e^{\varphi} = \frac{e^{\varphi_0}}{ 1 + (t/t_0)^2},
\nonumber\\
&& e^{\varphi_0}\rho_{0} = \frac{4}{t_0^2}= V_0 e^{2\varphi_0}.
\label{solf}
\end{eqnarray}

The evolution equations of the vector modes can be rederived by perturbing to first order 
Eq. (\ref{eq1}). In the case without sources, the result coincides in $D=4$, with Eqs. (\ref{v01}) and (\ref{v02}).
In the case with sources Eqs. (\ref{v01}) and (\ref{v02}) become
\begin{eqnarray}
&& \nabla^2V_{i} = -a^2 e^{\varphi} (p+\rho) {\cal V}_{i},
\label{source1}\\
&& V_{i}' = (\overline{\varphi}' + {\cal H}) V_i,
\label{source2}
\end{eqnarray}
where ${\cal V}_{i}$ is the fluctuation of the velocity field and where we used the explicit perturbed form of the energy--momentum tensor
\begin{equation}
\delta T_{i}^{0} = (p + \rho) {\cal V}_{i},\,\,\,\,\,\,\,\,\,\,\,\,\,u^{0} =\frac{1}{a},\,\,\,\,\,\,\,\,\,\,\,\,\delta u_{i} = a {\cal V}_{i}.
\end{equation}

Equations (\ref{pot1})--(\ref{dilaton1}) can be used in order to obtain the explicit solution for the gauge-invariant 
vector mode, which is
\begin{equation}
V_{i} = c_{i} e^{2\varphi_0} \frac{(\tau + \sqrt{\tau^2 +1})^{1/\sqrt{3}}}{\sqrt{\tau^2 +1}},
\label{Vf1}
\end{equation}
where, as before, $c_{i}$ is an integration constant carrying vector indices.
This shows that for $t\ll - t_0$,  $V_{i} \sim t^{- \frac{\sqrt{3} + 1}{\sqrt{3}}}$, as obtained previously. However,
for $t \gg t_0$ the dangerous growing mode turns into a decaying mode, i.e. $V_{i} \sim t^{( 1 - \sqrt{3})/\sqrt{3} }$. Furthermore, no divergence is present for $t\to 0$. 

In the case of the solution given in Eq. (\ref{solf}), Eqs. (\ref{source1}) and (\ref{source2}) lead to 
\begin{eqnarray}
&& V_{i} = c_{i} \frac{e^{\varphi_0} a_0}{1 + \tau^2}, 
\nonumber\\
&& {\cal V}_i = \frac{c_i k^2 }{a_0 \rho_0},
\label{Vf2}
\end{eqnarray}
where $k$ denotes the Fourier mode. 
\begin{figure}
\centerline{\epsfxsize = 9cm  \epsffile{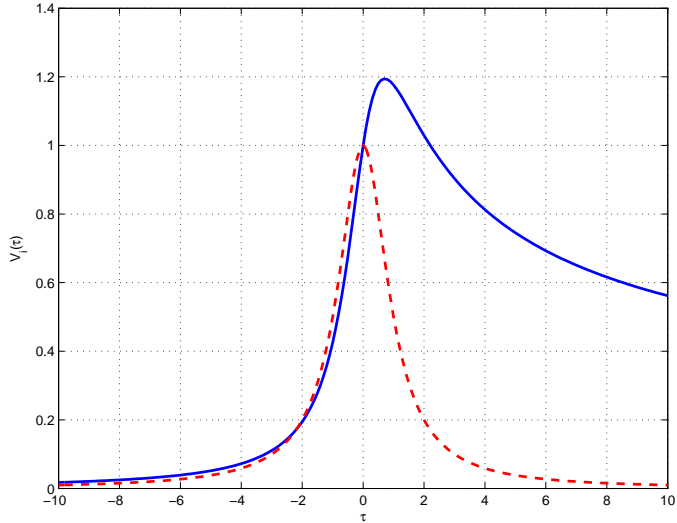}}
\vskip 3mm
\caption[a]{The evolution of the vector zero mode in the case without fluid sources (full line) and with fluid sources (dashed line). }
\label{figure1} 
\end{figure}
The time evolution of the gauge-invariant vector mode is illustrated in Fig. \ref{figure1} for the two results obtained in Eqs. (\ref{Vf1}) and (\ref{Vf2}).

The analysis reported so far shows that in some simple models of curvature bounce 
the growing mode in the vector sector matches with a decaying mode after the bounce.

\renewcommand{\theequation}{3.\arabic{equation}}
\section{Multidimensional curvature bounces}
\setcounter{equation}{0}
The analysis of gauge-invariant fluctuations in multidimensional and/or anisotropic cosmological models has
been discussed, through the years, in different contexts. One of the first attempts to adapt the original 
Bardeen \cite{bardeen} formalism to a multidimensional framework was discussed in \cite{ell} 
with the aim of performing a general discussion of the main features of cosmological fluctuations 
in Kaluza--Klein models \cite{kk}. Of related interests are some works discussing 
the theory of cosmological fluctuations in Bianchi models \cite{den,hu}.
In \cite{mg3}, a perturbative framework for the gauge-invariant discussion 
of multidimensional pre-big bang models was introduced. 

The problem of gauge-invariant fluctuations in models with extradimensions 
received new momentum in the context of brane models, and 
different approaches (see, for instance \cite{rob2,kolb2,gio1} and references therein) 
were developed in order to study the possible localization of 
modes of various spin  of both gravitational and non-gravitational origin.
In \cite{mg2} vector modes of the geometry were analysed in the case of a six-dimensional 
Abelian vortex leading to the localization of gravitational interaction (see also \cite{mg1} for a more 
general discussion of codimension-two braneworlds and of their vector excitations).

It is useful to notice, preliminarily, that the action given 
in Eq. (\ref{action1}) also admits classical solutions in the presence 
of internal (contracting) dimensions. In this section, the vector fluctuations of 
multidimensional cosmological models will be studied. The line element will be written, in this case, as
\begin{equation}
d s^2 = G_{\mu\nu} d x^{\nu} d x^{\mu} =  [dt^2 - a^2(t) g_{ij} dx^i d x^{j} ] - 
b^2(t) g_{AB} dy^{A} dy^{B}, 
\label{multidim}
\end{equation}
where $G_{\mu\nu}$ is the $D$-dimensional metric; $g_{ij}(x)$  and $g_{AB}(y)$ are 
the metric tensors of two maximally symmetric Euclidian manifolds.
The indices $\mu,\nu$ run over the full $D = d + n + 1$-dimensional
space-time; the $d$-expanding dimensions are typically three 
and the $n$-contracting  dimensions (labelled by capital Latin letters) will be, for a while, taken to be 
 generic but the case $n=1$ 
will be of particular interest in the following part of the investigation. 

From Eq. (\ref{multidim}) in the particular case $g_{ij} = \delta_{ij}$ and $g_{AB} = \delta_{AB}$, 
Eqs. (\ref{eq1}) and (\ref{eq2})
can be written in explicit terms by introducing the expansion rates $H=\dot{a}/a$ and $F =\dot{b}/b$ 
(the dot denotes a derivation with respect to the cosmic time coordinate):
\begin{eqnarray}
&& \dot{\overline{\varphi}}^2  - d H^2 - n F^2 - V= 0
\label{m1}\\
&& \dot{H} = H \dot{\overline{\varphi}},
\label{m2}\\
&&  \dot{F} = F \dot{\overline{\varphi}},
\label{m3}\\
&& 2 \ddot{\overline{\varphi}} - \dot{\overline{\varphi}}^2 - d H^2 - nF^2 + V - \frac{\partial V}{\partial\overline{\varphi}}=0.
\label{m4}
\end{eqnarray}
Equations (\ref{m1})--(\ref{m3})  come, respectively, from the $(0,0)$, $(i,j)$, $(a,b)$ components of Eq. (\ref{comb}), while Eq. (\ref{m4}) 
can be derived from Eq. (\ref{eq2}).  

In the case of $V= -V_0e^{4 \overline{\varphi}}$, Eqs. (\ref{m1})--(\ref{m4}) are solved by the following ansatz  
\begin{eqnarray}
&&a(t) = (\tau + \sqrt{\tau^2 + 1})^{\frac{1}{\sqrt{d + n}}},\,\,\,\,\,\,\,\,\, b(t) =  (\tau + \sqrt{\tau^2 + 1})^{-\frac{1}{\sqrt{d + n}}},
\label{solint1}\\
&&\overline{\varphi} = \varphi_0 - \frac{1}{2} \ln{[1 + \tau^2]}, \,\,\,\,\,\,\,\,\,\, \tau =\frac{t}{t_0}.
\label{solint2}
\end{eqnarray}
The solution given in Eqs. (\ref{solint1}) and (\ref{solint2}) describes 
the expansion of the $d$ dimensions and the contraction (at a dual rate) of the 
$n$ internal dimensions. 
 
Recalling the relation between the cosmic time coordinate $t$ and the conformal time coordinate $\eta$,
i.e. $a(\eta) d\eta = dt$, Eqs. (\ref{m1})--(\ref{m4}) can be written  as
\begin{eqnarray}
&& {\overline{\varphi}'}^2 - d {\cal H}^2 - n {\cal F}^2 - V a^2 =0,
\label{m1a}\\
&& {\cal H}' = {\cal H}^2 + {\cal H} \overline{\varphi}',
\label{m2a}\\
&&   {\cal F}' = {\cal H}{\cal F} + {\cal F} \overline{\varphi}', 
\label{m3a}\\
&& {\overline{\varphi}'}^2 - 2 \overline{\varphi}'' + 2 {\cal H} \overline{\varphi}' + 
d {\cal H}^2 + n {\cal F}^2 + 
\frac{\partial V}{\partial \overline{\varphi}} a^2 - V a^2 =0.
\label{m4a}
\end{eqnarray}
where, in full analogy with the four-dimensional case,  the rates of expansion
in conformal time have been defined:
\begin{equation} 
{\cal H} =(\ln{a})',\,\,\,\,\, {\cal F}= (\ln{b})', \,\,\,\,\, '= \frac{\partial}{\partial\eta}.
\end{equation}

\begin{figure}
\centerline{\epsfxsize = 9cm  \epsffile{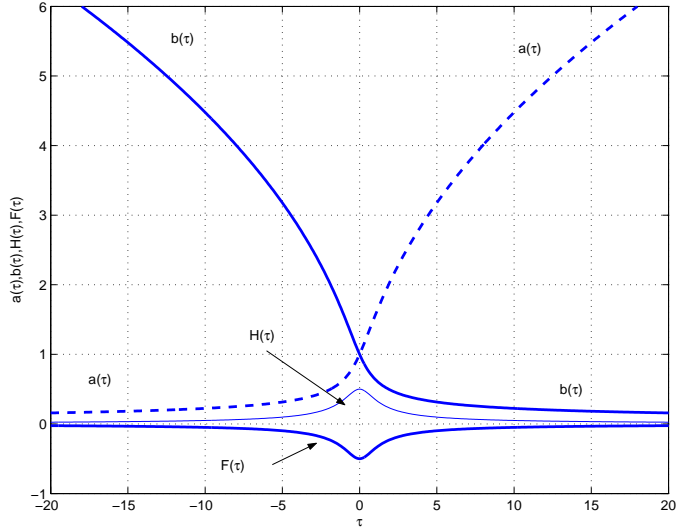}}
\vskip 3mm
\caption[a]{The time-evolution of the scale factors $a(\tau)$ (solid line) and $b(\tau)$ (dashed line) 
introduced in Eq. (\ref{explscale}). The time evolution of the associated 
rates of expansion and contraction is also reported.}
\label{figure2} 
\end{figure}

A particularly interesting case is the one where only one extradimension is present (i.e. $n =1$).
 In this case, setting $n=1$ and $d=3$ in Eq. (\ref{solint1}) the scale factors are 
\begin{equation}
a(t) = \frac{1}{b(t)} = \sqrt{\tau + \sqrt{\tau^2 +1}}.
\label{explscale}
\end{equation}
From Eq. (\ref{explscale}), it can be easily deduced that 
 the scale factor $a(t)$ smoothly interpolates between $a(t) \sim t^{-1/2}$ for $t\to -\infty$ and $a(t)\sim t^{1/2}$ 
for $t\to +\infty$. Moreover, $b(t)$ will contract like $t^{1/2}$ for $t\ll -t_0$ and will contract like 
$t^{-1/2}$ for $t\gg t_0$. Both the scale factors and the curvature invariants are regular 
in the origin and for $|t|\to \infty$. Finally, a pleasant feature of the case $d=3$ and $n=1$ is that 
in the limit $t\to +\infty$ the usual radiation-dominated evolution is correctly reproduced.
In Fig. \ref{figure2} the time evolution of the scale factors $a(\tau)$ and $b(\tau)$ and of the associated rates, i.e.  $H(\tau)$ and $F(\tau)$, 
is illustrated. This model describes the decoupling of internal and external dimensions.

In the following part of the present Section, the vector modes arising in the model of Eq. (\ref{explscale})
will be specifically analysed. The equations of motion will be derived in a gauge-independent way.
The fluctuations of a five-dimensional 
line element of the type of Eq. (\ref{multidim}) are, in general, described by $15$ degrees of freedom which can be formally 
decomposed in terms of their transformation properties with respect to three-dimensional rotations as
\begin{eqnarray}
&&\delta G_{00} = 2 a^2 \phi,
\nonumber\\
&& \delta G_{0 i} = - a^2 \partial_{i} P - a^2 Q_{i},
\nonumber\\
&& \delta G_{0 y} = - a b C,
\nonumber\\
&&\delta G_{i j} = 2 a^2 \psi \delta_{i j} - 2 a^2 \partial_{i} \partial_{j} E + a^2 h_{i j}
+ a^2 (\partial_{i} W_{j} + \partial_{j} W_{i}) ,
\nonumber\\
&& \delta G_{i y} = - a b \partial_{i} D - a b H_{i},
\nonumber\\
&& \delta G_{yy} = 2 b^2 \xi,
\end{eqnarray}
where 
\begin{eqnarray}
&& \partial_{j} h_{i}^{j}=0,~~~~h_{i}^{i} =0,
\label{tens}\\
&& \partial_{i} W^{i}=0,~~~~~~\partial_{i} Q^{i}=0,~~~~\partial_{i}H^{i} =0. 
\label{vec}
\end{eqnarray}
According to (\ref{tens})  $h_{ij}$ carries two degrees of freedom since it is symmetric in the two indices.
Equation (\ref{vec}) tells us  that  $W_{i}$, $Q_{i}$ and $H_{i}$ carry overall six degrees of freedom.  The remaining 
degrees of freedom are seven scalars, which will not be specifically discussed since the attention of the 
present analysis is mainly centred on the vector modes.

The  vectors $W_{i}$, $Q_{i}$ and $H_{i}$ obey a set of differential equations, which 
we ought to derive now. Notice, preliminarily, that in terms of the divergenceless gauge parameter $\zeta_i$  
the vectors  $W_{i}$, $Q_{i}$ and $H_{i}$ transform, for infinitesimal diffeomorphisms preserving the vector nature of the 
fluctuation, as  
\begin{eqnarray}
&& \tilde{Q}_{i} = Q_{i} - \zeta_{i}',
\nonumber\\
&& \tilde{W}_{i}= W_{i} + \zeta_{i},
\nonumber\\
&& \tilde{H}_{i} = H_{i} - \frac{a}{b} \partial_{y} \zeta_{i}, 
\end{eqnarray}
As in the four-dimensional case, gauge-invariant fluctuations can be defined, namely
\begin{eqnarray}
&& V_{i} = Q_{i} + W_{i}',
\label{gione}\\
&& Z_{i} = H_{i} + \frac{a}{b} \partial_{y} W_{i}.
\label{gitwo}
\end{eqnarray}
Notice that (\ref{gione}) corresponds to the gauge-invariant fluctuation of the four-dimensional case.

In order to derive the coupled evolution of $V_{i}$ and $Z_{i}$, Eq. (\ref{eq1}) shall be linearized 
to first order in $Q_{i}$, $W_{i}$ and $ H_{i}$. By inserting the definitions (\ref{gione}) and (\ref{gitwo}) into 
the obtained equations, the system will only contain the relevant gauge-invariant combinations, i.e. $V_{i}$ and $Z_{i}$.

From Eq. (\ref{eq1}), the general form of the various components of the perturbed equation can be written, in a compact form, as 
\begin{eqnarray}
&&  a^2 \delta {\cal G}_{i}^{j} - \frac{{\varphi}'}{2} [ \partial_{i} Q^{j} + \partial^{j} Q_{i}] -
\frac{{\varphi}'}{2} [ \partial_{i} {W^{j}}' + \partial^{j} W_{i}'] =0,
\label{p1}\\
&& a^2 \delta {\cal G}_{i}^{y} + \frac{a }{2 b} \varphi' [ H_{i}' + ({\cal F} - {\cal H}) H_{i}] 
 - \frac{\varphi'}{2} \frac{a^2}{b^2} \partial_{y} Q_{i} =0,
\label{p2}\\
&& a^2 \delta {\cal G}_{0}^{i} - Q^{i}( \varphi'' - 2{\cal H}\varphi') + \frac{a^2}{2} \frac{\partial V}{\partial \overline{\varphi}} Q^{i} =0.
\label{p3}
\end{eqnarray} 
In order to derive Eqs. (\ref{p1})--(\ref{p3}) the perturbed connections reported in the appendix (see Eq. (\ref{christoffel}))
have been inserted in the linearized form of Eq. (\ref{eq1}), bearing in mind that the 
fluctuation of the induced metric (\ref{induced}) implies
\begin{equation}
\delta \gamma^{i}_{0} = Q^{i}.
\end{equation}

From Eqs. (\ref{p1})--(\ref{p3}), using Eqs. (\ref{r0iup})--(\ref{riyup}) reported in the appendix, the following expressions are obtained:
\begin{eqnarray}
&& \frac{1}{2} \biggl[ (\partial_{i}Q^{j} + \partial^{j} Q_{i})' - ( \overline{ \varphi}' + {\cal H}) (\partial_{i} Q^{j} + \partial^{j}Q_{i}) \biggr]
- \frac{a}{2 b} \partial_{y}\biggl[ \partial_{i}H^{j} + \partial^{j} H_{i}\biggr] 
\nonumber\\
&&+ \frac{1}{2} \biggl[ (\partial_{i}W^{j} + \partial^{j} W_{i})'' + ( \overline{ \varphi}' + {\cal H}  ) (\partial_{i} W^{j} + \partial^{j}W_{i})' \biggr]
\nonumber\\
&&- \frac{a^2}{2 b^2} \partial_{y}\biggl[ \partial_{i}W^{j} + \partial^{j}W_{i}\biggr] =0,
\label{p1ex}\\
&& Q_{i} \biggl[{\overline{\varphi}'}^2  + 3 {\cal H}^2 + {\cal F}^2 - 2 \overline{\varphi}'' + 2 {\cal H} \overline{\varphi}' - Va^2 + \frac{\partial V}{\partial \overline{\varphi}} a^2 \biggr] 
\nonumber\\
&& +\frac{1}{2} \nabla^2 Q_{i} + \frac{1}{2} \nabla^2 W_{i}'  + \frac{a^2}{2 b^2} \nabla^2_{y} Q_{i} 
- \frac{a}{2 b} \partial_{y} \biggl[ H_{i}' + ({\cal F} - {\cal H}) H_{i} \biggr] =0,
\label{p2ex}\\
&& \frac{1}{2} \partial_{y} \biggl[ Q_{i}' + ( {\cal H} - 2{\cal F} - \overline{\varphi}') Q_{i}\biggr] + \frac{1}{2} \partial_{y} \nabla^2 W_{i} 
\nonumber\\
&&-
\frac{b}{2 a} \biggl[ H_{i}'' - (\overline{\varphi}' + {\cal H}) H_{i}' - \nabla^2 H_{i} - ({\cal H} - {\cal F})^2 H_{i} - \nabla^2 H_{i} \biggr] =0.
\label{p3ex}
\end{eqnarray}
Recalling the definitions of the gauge-invariant fluctuations given in Eqs. (\ref{gione}) and (\ref{gitwo}), 
Eq. (\ref{p1ex}) becomes
\begin{equation}
- \frac{a}{2 b} \partial_{y}[ \partial_{i} Z^{j} + \partial^{j}Z_{i}] + 
\frac{1}{2} \biggl[(\partial_{i} V^{j} + \partial^{j}V_{i})' - ({\overline{\varphi}}' + {\cal H}) 
(\partial_{i} V^{j} + \partial^{j}V_{i})\biggr] =0,
\label{ij}
\end{equation}

By virtue of Eq. (\ref{m4a}), the first term of Eq. (\ref{p2ex}) vanishes. 
The use of  the gauge-invariant fluctuations $V_{i}$ and $Z_{i}$ into Eq. (\ref{p2ex}) then leads to
\begin{equation}
- \frac{a}{2  b} \partial_{y}[ {Z_{i}}' + ({\cal F} - {\cal H}) Z_{i}] 
+ \frac{a^2}{2 b^2} \nabla^2_{y} V_{i} + \frac{1}{2} \nabla^2 V_{i}=0.
\label{0i}
\end{equation}
Finally, inserting Eqs. (\ref{gione}) and (\ref{gitwo}) into Eq. (\ref{p3ex}), the last gauge-invariant evolution equation will be:
the $(iy)$ equation
\begin{equation}
\frac{1}{2 } \partial_{y}[ V_{i}' + ( {\cal H} - 2 {\cal F} - \overline{{\varphi}}') V_{i}] 
- \frac{b}{2a} [ Z_{i}'' - ( {\overline{\varphi}}' + {\cal H} )Z_{i}' - ({\cal H} -{\cal F})^2  Z_{i} - \nabla^2 Z_{i} ] =0.
\label{iy}
\end{equation}

Equations (\ref{ij})--(\ref{iy}) can also be derived with gauge-dependent methods, as can be 
expected from general considerations. For instance, fixing the gauge in such away that $W_{i}=0$ the set of equations
(\ref{ij})--(\ref{iy}) can be  exactly reproduced, as  was checked by explicit calculation.
By  formally taking the limit $Z_{i}\to 0$ and by setting to zero the derivatives with respect to the internal coordinate $y$, 
 the four-dimensional equations derived in Eqs. (\ref{v01}) and (\ref{v02}) are formally 
recovered.  

\renewcommand{\theequation}{4.\arabic{equation}}
\section{Spectrum of vector modes}
\setcounter{equation}{0}

For the purposes of this investigation it is interesting to consider the spectrum of the 
zero modes with respect to the internal dimensions. This is achieved by setting to zero 
all the terms containing one (or more ) derivatives with respect to $y$.
Thus, from Eqs. (\ref{0i})--(\ref{iy}) the following system:
\begin{eqnarray}
&&\nabla^2 Q_{i} =0, 
\label{I}\\
&& Q_{i}' = ({\cal H} +\overline{\varphi}') Q_{i},
\label{II}\\
&& H_{i}'' - ( {\cal H} +\overline{\varphi}') H_{i}' - ( {\cal H} - {\cal F})^2 H_{i} - \nabla^2H_{i} =0,
\label{III}
\end{eqnarray}
 is obtained.
The evolution of $H_{i}$ and $Q_{i}$ is, in this approximation, decoupled.  Furthermore, it can be 
noticed that the evolution of $Q_{i}$ is the one already discussed in the four-dimensional case, 
with the minor difference that, in the present case, the background evolution differs from the
four-dimensional case. 
From Eq. (\ref{II}) we then have 
\begin{equation}
Q_{i} (t)= c_i 
\frac{ \sqrt{\tau (\tau^2 + 1) + (\tau^2 +1)^{3/2}}}{\tau^2 + 1}, 
\end{equation}
which also implies that $Q_{i} \sim t^{-1/2}$ for $ t \gg t_0$. Hence, even if
the time dependence is different from the four-dimensional case, it is still 
true that the $Q_{i}$ decreases at large (positive) times even if 
 the solution is growing just before the bounce.

Let us now concentrate on the evolution of $H_{i}$ whose equation can be written, in Fourier 
space and for a single vector polarization as\footnote{Notice that in the following the index denoting 
the Fourier mode will be omitted in order to avoid confusion with the vector polarization.}
\begin{eqnarray}
&&u''  + [k^2 - {\cal U}(\eta)] u =0,
\label{u}\\
&& {\cal U}(\eta) = ({\cal H} - {\cal F})^2 - \biggl( \frac{e^{- 
\overline{\varphi}/2}}{\sqrt{a}}\biggr)'' \sqrt{a} e^{\overline{\varphi}/2},
\label{potU}
\end{eqnarray}
where $H_{i} = \sqrt{a} e^{\overline{\varphi}/2}u_{i}$. 
This equation will be numerically solved in a moment  for different values of $k$. However, in order 
to cross-check the numerical results it is useful to anticipate the analytical estimate. For this purpose a useful observation is
 that, for times that are sufficiently far from the 
bounce, the asymptotic solutions are an excellent approximation for the background evolution.

With this logic in mind, for  $t \ll - t_{0}$ the solution given by Eqs. (\ref{solint1}) and 
(\ref{solint2}) can be approximated, in conformal time, by
\begin{equation}
a(\eta) \sim (-\eta)^{-1/3},\,\,\,\,\,\,\,\,\,\,b(\eta) \sim (-\eta)^{1/3},
\,\,\,\,\,\,\,\,\overline{\varphi} \sim - \frac{2}{3} \ln{(-\eta)}.
\label{tminus}
\end{equation}
For $t \gg  t_{0}$, the solution (\ref{solint1}) and (\ref{solint2}) behaves, in conformal time, as
\begin{equation}
a(\eta) \sim \eta ,\,\,\,\,\,\,\,\,\,\,b(\eta) \sim \frac{1}{\eta},
\,\,\,\,\,\,\,\,\overline{\varphi} \sim - 2 \ln{\eta}.
\label{tplus}
\end{equation}
Therefore, in the two opposite limits, Eq. (\ref{u}) is given, respectively, 
by  
\begin{eqnarray}
&&u_{i}'' + \biggl[k^2 - \frac{4 \nu^2 -1}{4\eta^2}\biggr] u_{i}= 0, 
\,\,\,\,\,\,\,\,\,\, \eta < - \eta_{1},
\label{nu}\\
&&u_{i}'' + \biggl[k^2 - \frac{4 \mu^2 -1}{4\eta^2}\biggr] u_{i}= 0, 
\,\,\,\,\,\,\,\,\,\, \eta >  \eta_{1},
\label{mu}
\end{eqnarray}
where $\eta_1 \sim \eta_0$ is a time scale comparable with the size of the bounce.

Inserting Eqs. (\ref{tminus}) and (\ref{tplus}) into Eq. (\ref{potU}) the indices $\mu$ and $\nu$ are determined to be 
 $\nu = 2/3$ and $\mu= 2$. 
Equations (\ref{nu}) and (\ref{mu}) 
can be solved exactly in the two limits. 
Now, imposing quantum-mechanical 
initial conditions for $\eta\to - \infty$,
\begin{equation}
 u_{-}(\eta) = \frac{\sqrt{-\pi\eta}}{2} e^{ i \frac{\pi}{2} ( \nu + 1/2)} H_{\nu}^{(1)}( - k\eta) ,~~~~~~~~~~\eta < -\eta_{0},
\label{uminus}
\end{equation}
the solution for $\eta \to +\infty$ can be written in terms of the 
mixing coefficients $c_{\pm}(k)$ 
\begin{equation}
u_{ +}(\eta) = \frac{\sqrt{\pi\eta}}{2} e^{ i \frac{\pi}{2} ( \mu + 1/2)}\biggl[ c_{-}(k)
H_{\mu}^{(1)}( k\eta) + c_{+}(k) e^{- i \pi (\mu +1/2)} H_{\mu}^{(2)}(k\eta) \biggr] ,~\eta > \eta_{0},
\label{uplus}
\end{equation}
 where $H^{(1)}_{\alpha}(z) = {H^{(2)}}^{\ast}_{\alpha}(z)= J_{\alpha}(z)+ i Y_{\alpha}(z)$ 
 are the usual Hankel functions of argument $z$ and index $\alpha$ \cite{magnus}. Notice 
that the phases in Eqs. (\ref{uminus}) and (\ref{uplus}) have been carefully selected in such a way as 
to recover mode functions with canonical quantum mechanical normalization in the limits $\eta\to \pm \infty$.
 
The two solutions (\ref{uminus}) and (\ref{uplus}) can be continuously matched across the bounce, i.e. imposing $u_{-} (-\eta_1) = u_{+}(\eta_1)$ 
and similarly for the first derivative.  The mixing coefficients turn out to be  
\begin{eqnarray}
&& c_{+}(k) = i \frac{\pi}{4} x_1 e^{i\pi( \nu +\mu + 1)/2} \biggl[ - \frac{\nu + \mu +1}{x_1} H_{\mu}^{(1)}(x_1) H_{\nu}^{(1)}(x_1) 
\nonumber\\
&&+
H_{\mu}^{(1)}(x_1) H_{\nu+ 1}^{(1)} (x_1) + H_{\mu+1}^{(1)}(x_1) H_{\nu}^{(1)}(x_1) \biggr],
\label{cpp}\\
&& c_{-}(k)=  i \frac{\pi}{4} x_1 e^{i\pi( \nu -\mu)/2} \biggl[ - \frac{\nu + \mu +1}{x_1} H_{\mu}^{(2)}(x_1) H_{\nu}^{(1)}(x_1) 
\nonumber\\
&& + H_{\mu}^{(2)}(x_1) H_{\nu+ 1}^{(1)}(x_1) + H_{\mu + 1}^{(2)}(x_1) H_{\nu}^{(1)}(x_1) \biggr],
\label{cpm}
\end{eqnarray}
satisfying the exact Wronskian normalization condition $|c_{+}(k)|^2 - |c_{-}(k)|^2 =1$.
In the small argument limit, i.e. $k\eta_1 \sim k\eta_0 \ll 1$ the leading term in Eq. (\ref{cpm}) leads to 
\begin{equation}
c_{-}(k) \simeq \frac{i~ 2^{\mu +\nu}}{4\pi} e^{i \pi(\nu - \mu)/2} x_1^{- \mu - \nu}  (\nu + \mu -1) \Gamma(\mu) \Gamma(\nu),
\label{cminanal}
\end{equation}
which implies, after inserting the specific values of $\mu$ and $\nu$ 
that the number of produced vector quanta is given, for each polarization, by 
\begin{equation}
|c_{-}(k)|^2 \sim |k/k_1|^{-2\gamma},\,\,\,\,\, \gamma = (\mu +\nu) \sim 2.6,
\label{pred}
\end{equation}
implying, in turn, that the energy density per logarithmic
frequency interval, i.e. $k^4 |c_{-}(k)|^2$,
increases at large distance-scales (small $k$)  as $ k^{- 1.12}$.

The same estimate will now be obtained through a totally different procedure, where the 
evolution equation  for $H_{i}$ is solved directly using the exact background solutions. 
Since the background solutions have a  simple form in cosmic time, it is more practical 
to solve the evolution equation of the fluctuations directly in cosmic time. This will 
not affect the determination of the mixing coefficients, which are independent 
of the specific parametrization of the time coordinate.

Equation (\ref{III}) becomes, in cosmic time, 
\begin{equation}
\ddot{H}_{i} - \dot{\overline{\varphi}} \dot{H}_{i}  +[ \omega^2 - (H- F)^2] H_{i} =0,
\label{cost1}
\end{equation}
where $\omega(t) = k/a(t)$ is the physical momentum.
By defining now the rescaled variable $v_{i} = e^{- \overline{\varphi}/2} H_{i}$, Eq. (\ref{cost1}) 
becomes:
\begin{equation}
\ddot{v}_{i} + \biggl[ \omega^2 + \frac{\ddot{\overline{\varphi}}}{2} - \frac{\dot{\overline{\varphi}}^2}{2 } - (H -F)^2 \biggr]v_{i}= 0.
\label{cost2}
\end{equation}
Inserting Eqs. (\ref{solint1}) and (\ref{solint2}) into Eq. (\ref{cost2}), the following simple equation  for a single 
vector polarization can be obtained:
\begin{equation}
\frac{d^2 v}{d\tau^2} + \biggl[ \frac{\kappa^2}{\tau \sqrt{\tau^2 + 1}} - \frac{ 3 (\tau^2 +2)}{4(\tau^2 + 1)^2}\biggr] v=0,
\label{cost3}
\end{equation}
where $\tau = t/t_0$ is the usual rescaled cosmic time and $\kappa = k t_0$.

\begin{figure}
\centerline{\epsfxsize = 9cm  \epsffile{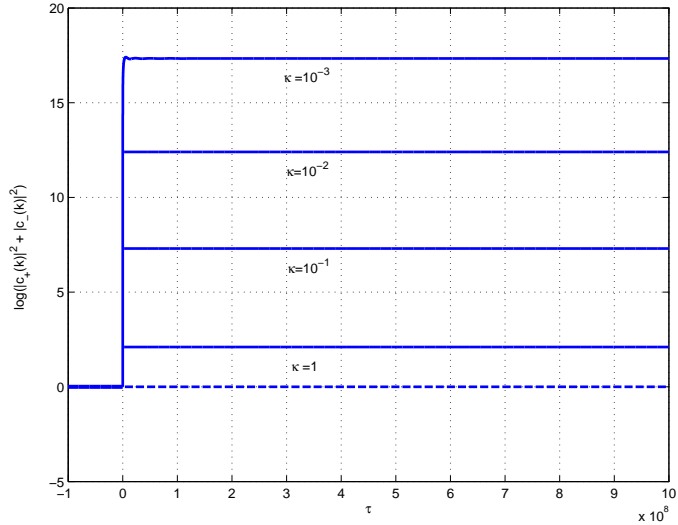}}
\vskip 3mm
\caption[a]{The numerical calculation of the mixing coefficients for different values of $\kappa = k t_0$, the 
comoving momentum rescaled through the size of the bounce. The dashed line is Eq. (\ref{diff}) whose right-hand side 
has been numerically computed. }
\label{figure3} 
\end{figure}

Notice that the variable $v$ defined in Eq. (\ref{cost3}) is simply related to the variable $u$ defined in Eq. (\ref{u}):
\begin{equation}
H_{i} = \sqrt{a} e^{\overline{\varphi}/2} u_{i} =  e^{\overline{\varphi}/2} v_{i}.
\end{equation}
Since $ v = \sqrt{a} u$, the quantum-mechanical initial 
conditions will be, in terms of $v$
\begin{equation}
v(t) = \frac{1}{\sqrt{2\omega}} e^{- i \int \omega dt},
\end{equation}
for $t \to -\infty$. For $t\to +\infty$ the solution can again be written in terms of the mixing coefficients 
$c_{\pm}$ as 
\begin{equation}
v(t) = \frac{1}{\sqrt{2 \omega}}\biggl[ c_{+}(k) e^{- i\int\omega dt } +c_{-}(k) e^{ i\int\omega dt } \biggr].
\end{equation} 
Equation (\ref{cost3}) was solved for different values of $k$ (or $\kappa =k t_0$, the dimensionless comoving 
momentum). Quantum-mechanical initial conditions are given when the given mode is 
inside the horizon at the initial time of integration. 
From the numerical solution for the real and imaginary parts of $v$ and $\dot{v}$, i.e.
\begin{equation}
v= f + i q,\,\,\,\,\,\,\,\,\,\, \dot{v} = g + i p,
\end{equation}
it is then easy to 
obtain the mixing coefficients after the relevant mode re-entered during radiation, i.e. 
\begin{eqnarray}
&&|c_{+}(k)|^2 + |c_{-}(k)|^2 =\frac{1}{\omega} \biggl[ \omega^2( f^2 + q^2) + \biggl(g - \frac{H}{2}f\biggr)^2 +\biggl(p - \frac{H}{2}q\biggr)^2 \biggr],
\label{sum}\\
&& |c_{+}(k)|^2 - |c_{-}(k)|^2 = 2 ( q g - f p). 
\label{diff}
\end{eqnarray}
The results of this calculation are reported in Fig. \ref{figure3}.
Recall that, according to Eq. (\ref{diff}),  $|c_{+}(k)|^2 - |c_{-}(k)|^2=1$ since $|c_{+}(k)|^2 - |c_{-}(k)|^2$ is proportional 
to the Wronskian of the solution.
From Fig. \ref{figure3} it is apparent that for an increase of one order of magnitude in $\kappa$, $|c_{-}(k)|^2$ increases by 
$5.1$ orders of magnitude, exactly as predicted in Eq. (\ref{pred}).

\renewcommand{\theequation}{5.\arabic{equation}}
\section{Concluding remarks}
\setcounter{equation}{0}

In the present paper the evolution of vector modes of the 
geometry has been discussed in the case of bouncing cosmologies 
both in four and in higher dimensions. The main suggestions 
following from a series of detailed examples can be summarized as follows:
\begin{itemize}
\item{} in the four-dimensional case, as also noted in \cite{BB}, 
a growing vector mode is present in the regime preceding the 
curvature bounce;
\item{} however, in the simple analytical models described 
in the present paper, the growing mode matches with a decaying 
mode after the curvature bounce, so that no further dangerous divergences 
are introduced;
\item{} in higher dimensions, more vector modes are expected and their 
spectrum was computed in a simple five-dimensional model of curvature bounce, 
where the compactification of the internal dimension is achieved dynamically;
\item{} the typical spectra of vector modes obtained in the specific (but 
fully consistent) example are red after the modes reenter the horizon.
\end{itemize}
The most problematic aspect of this analysis is represented by the possible 
occurrence of red spectra, which may be rather dangerous in models 
based on curvature bounces. 
\newpage

\begin{appendix}
\renewcommand{\theequation}{A.\arabic{equation}}
\setcounter{equation}{0}
\section{Vector modes of the geometry} 
The three divergenceless vectors appearing in five dimensions are defined as 
\begin{eqnarray}
&& \delta G_{0 i} = - a^2 Q_{i},
\nonumber\\
&& \delta G_{i j} = a^2 ( \partial_{i} W_{j} + \partial_{j} W_{i}),
\nonumber\\
&& \delta G_{i y}= - a b H_{i}
\label{v1}
\end{eqnarray}
and 
\begin{eqnarray}
&& \delta G^{0 i} = - \frac{ Q^{i}}{a^2},
\nonumber\\
&& \delta G^{i j} = -\frac{1}{a^2}  ( \partial^{i} W^{j} + \partial^{j} W^{i}),
\nonumber\\
&& \delta G^{i y} =  \frac{H^{i}}{a b}. 
\label{v2}
\end{eqnarray}
Notice that Eqs. (\ref{v1}) and (\ref{v2}) are general in the sense that no specific gauge choice has been 
imposed. 
From Eqs. (\ref{v1}) and (\ref{v2})
\begin{eqnarray}
&&\delta \Gamma_{i0}^{0} = {\cal H} Q_{i},
\nonumber\\
&& \delta \Gamma_{ij}^{0} = -\frac{1}{2} ( \partial_{i} Q_{j} + 
\partial_{j} Q_{i}) - {\cal H} (\partial_{i} W_{j} + \partial_{j} W_{i}) - \frac{1}{2}(\partial_{i} W_{j}' + \partial_{j} W_{I}') ,
\nonumber\\
&& \delta \Gamma_{00}^{i} = { Q^{i}}' + {\cal H} Q^{i},
\nonumber\\
&& \delta \Gamma_{i0}^{j} = \frac{1}{2}( \partial_{i} Q^{j} - \partial^{j} Q_{i}) - \frac{1}{2}(\partial^{j} W_{i}' + \partial_{i}{W^{j}}'),
\nonumber\\
&& \delta\Gamma_{y y}^{i} = \frac{b}{a} \partial_{y} H^{i} - \frac{b^2}{a^2} {\cal F} Q^{i} 
\nonumber\\ 
&& \delta \Gamma_{y i}^{0} = - \frac{1}{2} \partial_{y} Q_{i} + \frac{b}{2 a} [ H_{i}' + 
({\cal H} + {\cal F})H_{i}] ,
\nonumber\\
&& \delta \Gamma_{i0}^{y} = \frac{a}{2 b} [ H_{i}' + ({\cal F} - {\cal H}) H_{i}] - 
\frac{a^2}{2 b^2} \partial_{y} Q_{i} 
\nonumber\\
&& \delta \Gamma_{0 y}^{i} = \frac{1}{2} \partial_{y} Q^{i} + \frac{b}{2 a} [ {H^{i}}' + 
({\cal H} - {\cal F}) H^{i}],
\nonumber\\
&& \delta \Gamma_{i j}^{y} = \frac{a}{2 b} ( \partial_{i} H_{j} + \partial_{j} H_{i}) + \frac{a^2}{2 b^2}( \partial_{i} W_{j} + \partial_{j} W_{i}), 
\nonumber\\
&& \delta \Gamma_{i j}^{k} = - {\cal H} Q^{k} \delta_{i j} + \frac{1}{2} \partial^{k}(\partial_{i} W_{j} + \partial_{j} W_{i}) - 
\frac{1}{2}\partial_{j}( \partial^{k} W_{i} + \partial_{i} W^{k}) - \frac{1}{2} \partial_{i}(\partial_{j}W^{k} + \partial^{k}W_{j}),
\nonumber\\
&& \delta \Gamma_{i y}^{j} = \frac{b}{2 a} ( \partial_{i} H^{j} - \partial^{j} H_{i}) - \frac{1}{2} \partial_{y} (\partial^{j}W_{i} + \partial_{i}W^{j}).
\label{christoffel}
\end{eqnarray}
From the expressions of the Christoffel connections, the first-order fluctuations of the Ricci tensors can be determined:
\begin{eqnarray}
&& \delta R_{0 i} = [ {\cal H}' + 2 {\cal H}^2 + {\cal H} {\cal F}] Q^{i} -
\frac{a^2}{2 b^2} \nabla^2_{y} Q_{i} - \frac{1}{2} \nabla^2 Q_{i}  - \frac{1}{2} \nabla^2 W_{i}'
\nonumber\\
&& +
\frac{a}{2 b} \partial_{y} [ H_{i}' + ({\cal F} - {\cal H}) H_{i}],
\label{r0i}\\
&& \delta R_{i j} = \frac{a}{2 b} \partial_{y} [ \partial_{i} H_{j} + \partial_{j} H_{i}] 
- \frac{1}{2} \biggl\{ ( \partial_{i} Q_{j} + \partial_{j} Q_{i})' + ( 2 {\cal H} + {\cal F})   ( \partial_{i} Q_{j} + \partial_{j} Q_{i})\biggr\}
\nonumber\\
&&- \frac{1}{2}( \partial_{i}W_{j}'' + \partial_{j}W_{i}'') - \frac{1}{2}(2 {\cal H} + {\cal F}) (\partial_{i}W_{j}' + \partial_{j}W_{i}')
\nonumber\\
&&+ \frac{a^2}{2 b^2} \partial_{y} ( \partial_{i} W_{j} + \partial_{j}W_{i}) - \frac{1}{2} (\partial_{i}W_{j} + \partial_{j}W_{i}) 
( 2 {\cal H}' + 4 {\cal H}^2 + 2 {\cal H} {\cal F}) ,
\label{rij}\\
&& \delta R_{i y} = - \frac{1}{2} \partial_{y}[ Q_{i}' + (4 {\cal H} - {\cal F}) Q_{i}] - \frac{1}{2}\partial_{y} \nabla^2 W_{i}
\nonumber\\
&& +
\frac{b}{2 a} \biggl\{ H_{i}'' + (2 {\cal H} + {\cal F}) H_{i}' + ( {\cal H}' + {\cal F}' + 
{\cal H}^2 + 5 {\cal H} {\cal F}) H_{i} - \nabla^2 H_{i}\biggr\}.
\label{riy}
\end{eqnarray}
In Eqs. (\ref{r0i})--(\ref{riy}), $\nabla^2_{y}$ denotes the Laplacian of the internal dimensions.
The perturbed Ricci tensors  with mixed indices are also useful for a swifter calculation 
of the perturbed equations: 
\begin{eqnarray}
&& \delta R_{0}^{i} = \frac{Q^{i}}{a^2}[ 2 {\cal H}' + {\cal F}' - 2 {\cal H}^2 
- 2 {\cal H}{\cal F} + {\cal F}^2] + \frac{1}{2 b^2} \nabla^2_{y} Q^{i} +\frac{1}{2 a^2 } \nabla^2 Q^{i}
\nonumber\\
&&+ \frac{1}{2a^2} \nabla^2 W_{i}
 - \frac{1}{2 a b} \partial_{y}[ {H^{i}}' + 
({\cal F} - {\cal H}) H^{i}],
\label{r0iup}\\
&& \delta R_{i}^{j} = - \frac{1}{2 a b} \partial_{y} [ \partial_{i} H^{j} -\partial^{j} H_{i}]
+ \frac{1}{2 a^2} \biggl\{ (\partial_{i} Q^{j} + \partial^{j} Q_{i})' + 
( 2 {\cal H} + {\cal F})  (\partial_{i} Q^{j} + \partial^{j} Q_{i})\biggr\}
\nonumber\\
&& + \frac{1}{2 a^2} \biggl\{ (\partial_{i} W^{j} + \partial^{j} W_{i})'' + 
( 2 {\cal H} + {\cal F})  (\partial_{i} W^{j} + \partial^{j} W_{i})'\biggr\}
- \frac{1}{2b^2}\partial_{y}[ \partial_{i}W^{j} + \partial^{j}W_{i}],
\label{rijup}\\
&& \delta R_{i}^{y} = \frac{1}{2 b^2} \partial_{y}[ Q_{i}' + (4 {\cal H} - {\cal F}) Q_{i}] +\frac{1}{2b^2} \partial_{y} \nabla^2 W_{i}
\nonumber\\
&&- \frac{1}{2 a b} \biggl\{ H_{i}'' + (2 {\cal H} + {\cal F}) H_{i}' + 
( {\cal F}' - {\cal H}' - 3 {\cal H}^2 + 3 {\cal H}{\cal F})H_{i} - \nabla^2 H_{i} \biggr\}.
\label{riyup}
\end{eqnarray}
Clearly, since vector fluctuations are parametrized by divergence-less vectors in three 
dimensions, their contribution to the first order fluctuation of the Ricci scalar vanishes. Hence, recalling the 
definition of the Einstein tensor ${\cal G}_{\mu}^{\nu}$, 
\begin{equation}
\delta {\cal G}_{0}^{i}= \delta R_{0}^{i},\,\,\,\,\,\,\,\delta {\cal G}_{i}^{j}= \delta R_{i}^{j},\,\,\,\,\,\,\,
\delta {\cal G}_{i}^{y} = \delta R_{i}^{y}.
\end{equation}
\end{appendix}

\end{document}